\documentclass[aps,preprint,showkeys,a4paper,tightenlines,superscriptaddress,12pt]{revtex4}
\usepackage{eurosym}
\usepackage{amsfonts}
\usepackage{amsmath}
\usepackage{amssymb}
\usepackage{graphicx}
\usepackage{subfig}
\usepackage{caption}
\usepackage{setspace}

\providecommand{\U}[1]{\protect\rule{.1in}{.1in}}

\begin{document}

\section*{SUBMITTED TO "MECHANICS OF MATERIALS" ON 4th OCTOBER 2020}

\title{Rate-dependent adhesion of viscoelastic contacts. Part II:
numerical model and hysteresis dissipation}
\author{G. Violano}
\affiliation{Department of Mechanics, Mathematics and Management, Polytechnic University
of Bari, Via E. Orabona, 4, 70125, Bari, Italy}
\author{A. Chateauminois}
\affiliation{Soft Matter Science and Engineering Laboratory (SIMM), PSL Research
University, UPMC Univ Paris 06, Sorbonne Universit\~{A}\copyright s, ESPCI
Paris, CNRS, 10 rue Vauquelin, 75231 Paris cedex 05, France}
\author{L. Afferrante}
\email{guido.violano@poliba.it}
\affiliation{Department of Mechanics, Mathematics and Management, Polytechnic University
of Bari, Via E. Orabona, 4, 70125, Bari, Italy}

\begin{abstract}
In this paper, we propose a numerical model to describe the adhesive normal
contact between a `rigid' spherical indenter and a viscoelastic rough
substrate. The model accounts for dissipative process under the assumption
that viscoelastic losses are localized at the (micro)-contact lines.
Numerical predictions are then compared with experimental measurements,
which show a strong adhesion hysteresis mostly due to viscous energy
dissipation occurring during pull-off. This hysteresis is satisfactorily
described by the contact model which allows to distinguish the energy loss
due to material dissipation from the adhesion hysteresis due to elastic
instability.

Our analysis shows that the pull-off force required to detach the surfaces
is strongly influenced by the detachment rate and the rms roughness
amplitude, but it is almost unaffected by the maximum load from which
unloading starts. Moreover, the increase in the boundary line separating
contact and non-contact regions, observed when moving from smooth to rough
contacts, negligibly affects the viscous dissipation. Such increase is much
less significant than the reduction in contact area, which therefore is the
main parameter governing the strong decrease in the effective surface energy.
\end{abstract}

\maketitle

\section{Introduction}

Fuller \& Tabor \cite{FT} firstly showed that the pull-off force, i.e. the
tensile load required to detach two contacting bodies, is strongly reduced
when the surface roughness is increased. More recently, Persson \& Tosatti 
\cite{PT2001} found that adhesion leads to an increase in the real contact
area, even when no pull-off force is detected.

Adhesive interactions are predominant at the nanometer scale for bodies and
systems with a high surface to volume ratio \cite{vakis2018}. However,
adhesion is still observed at macroscopic scales when the contacting bodies
are soft. Typical examples are pressure sensitive adhesives (PSA) \cite{PSA}%
, soft rubbers \cite{Barthel2009}, sthrechable electronics \cite{electronics}%
, and biomimetic devices \cite{biomimetic}.

In several experiments \cite{MB1978,tiwary2017}, the detachment behaviour of
soft matter is found to be rate-dependent as a result of the intrinsic
viscoelasticity of the material. In such conditions, the effective work of
adhesion $\Delta \gamma _{\mathrm{eff}}$ may be strongly increased compared
to the quasi-static value $\Delta \gamma _{\mathrm{0}}$.

Usually, the contact between soft media is described by the classical JKR
adhesion theory \cite{JKR}. However, the JKR theory applies for purely
elastic media as it neglects rate-dependence.

The coexistence of adhesion, viscoelasticity and surface roughness has been
experimentally investigated in numerous works \cite%
{tiwary2017,Lorenz2013,martina2012,dorogin2017,dorogin2018}. In such works,
loading-unloading experimental curves are usually fitted by exploiting the
JKR\ theory, with the stratagem of using different values of the work of
adhesion and elastic modulus for the loading and unloading phase,
respectively.

In literature, there is a lack of analytical and numerical models aimed at
describing the adhesive contact of viscoelastic bodies in presence of
surface roughness. Haiat \& Barthel (HB) \cite{Haiat2007} proposed an
approximate model for the contact of viscoelastic rough surfaces based on
the Greenwood \& Williamson (GW) model \cite{GW1966}. From an experimental
perspective, elucidation and validation of these models using microscopic
randomly rough surfaces such as abraded or bead blasted surfaces is
compromised by the difficulties in the measurement of the actual
distribution of micro-contact areas at the micrometer scale.

In Ref. \cite{Parte1}, we experimentally studied the detachment of a rigid
indenter from soft PolyDiMethylSiloxane (PDMS) substrates with smooth and
rough surface. Specifically, roughness was obtained by texturing the surface
with spherical identical micro-asperities with controlled height and spatial
distributions. The designed patterned surfaces allow for a precise determination of the real contact area from micro-contact visualization. Interestingly, we found simple scaling laws relating the
contact radius $a$ and the contact line velocity $v_{\mathrm{c}}$ measured
at the macro and microscale. 

Moving from the observed similarity between macro and microscale contacts,  we develop a numerical model able to describe both the
loading and unloading phases occurring in typical JKR tests on rough
samples. Specifically, the model exploits a discrete version of the Fuller
\& Tabor (FT) multiasperity model \cite{FT} to simulate the loading phase.
The unloading phase is instead modelled on the basis of the solution
proposed by Muller \cite{Muller1999}, with the assumption that the
parameters of Muller's model, which are experimentally identified at the
macroscale, can be applied to the microscale as they correspond to similar
contact line velocities $v_{\mathrm{c}}$.

\section{Experimental set-up}

For details about the experimental setup, the manufacturing of the PDMS
samples and the experimental procedure used during indentation tests we
refer the reader to Ref. \cite{Parte1}. Here, we simply summarize some
aspects regarding the generated patterned surfaces.

Roughness on the top of PDMS\ samples is obtained by texturing them with
identical spherical microasperities with controlled height and spatial
distributions. Patterned surfaces were obtained by moulding PDMS in
Poly(MethylMethAcrylate) (PMMA) micro-milled forms using ball-end mills with
a radius of $100$ $\mathrm{\mu m}$. In order to reduce the microscale
roughness induced by the milling process and thus to enhance adhesion, the
spherical cavities of the PMMA molds have been exposed to a saturated $%
\mathrm{CHCl3}$ vapor for $30$ minutes. As a result of surface
plasticization of the glassy acrylate polymer, surface tension effects were
previously found to result in a smoothening of the surface of the spherical
cavities of the mold \cite{ACITO2019}. As a consequence, an increase in the
radius of the spherical bumps is observed up to a $10\%$ of the nominal
value.

The size of these micro-asperities allows for an optical detection of the
individual micro-contact areas, which in turn provides the relationship
between the real contact area $A$ and the applied normal force $F$.

The patterned surfaces are generated with a squared nominal area of $10$ $%
\mathrm{mm}^{2}$, where asperities are randomly distributed with a density
of $2\times 10^{7}$ $\mathrm{m}^{-2}$. Asperities are collocated with a
non-overlapping constraint.

The first pattern is a regular square network of spherical caps having all
the same height of $40$ $\mathrm{\mu }$\textrm{m}; the other patterns are
characterized by spherical caps with heights distributed according to
Gaussian distributions with standard deviations $\sigma =5$ $\mathrm{\mu }$%
\textrm{m} and $\sigma =10$ $\mathrm{\mu }$\textrm{m}, respectively.

Indentation experiments of the glass lens on the smooth part of the PDMS
sample were performed at increasing imposed loads and the contact radius was
measured at each load step after the achievement of adhesive equilibrium.
The resulting contact radius vs. load data were fitted according to the JKR
theory. The fit allows to evaluate the values of the reduced elastic modulus
($E^{\ast }=0.83$ $\mathrm{MPa}$) and adhesion energy ($\Delta \gamma
_{0}=0.037$ $\mathrm{J/m}^{2}$).

The experimental tests on the rough patterns are performed in $6$ different
locations for each pattern in order to have $6$ realizations of the surface
topography.

During loading, contact tests are performed under fixed load conditions. The
applied load is incremented step by step, with an incremental step equal to $%
4$ \textrm{mN}. Once each load step is reached, contact is maintained for a
long time ($800$ \textrm{s}). In such conditions, we are sure that adhesive
equilibrium is reached as viscoelastic effects are totally dissipated \cite{ACITO2019}.

Unloading tests are instead performed by fixing the driving velocity. In such case, viscous effects occur during the detachment process
and JKR theory is no longer valid.

\section{Numerical model}

\begin{figure}[tbp]
\begin{center}
\includegraphics[width=15.0cm]{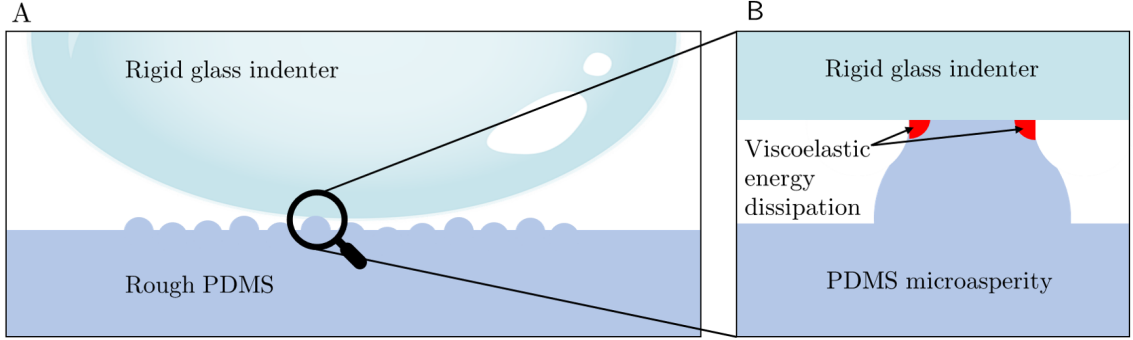}
\end{center}
\caption{A) The problem under investigations: the normal adhesive contact
between a rigid smooth spherical lens and soft rough PDMS. B) Detachment of
a viscoelastic micro-asperity.}
\label{modello}
\end{figure}

Fig. \ref{modello}A shows the frictionless adhesive contact between a
`rigid' indenter and a soft substrate of PDMS material textured with
spherical micro-asperities of identical radius of curvature.

As above explained, the loading process was executed experimentally under
conditions of adhesive equilibrium. Viscoelasticity is hence negligible and
micro-asperities were modeled as elastic spheres in the simulation.
Conversely, viscous dissipation is no longer negligible during unloading,
(fig. \ref{modello}B). Here, the problem is treated under the assumption
that viscoelastic losses are restricted to the contact line while the bulk
of the micro-contact zone behaves elastically.

This hypothesis is supported by the low glass transition temperature of the
used silicone ($T_{g}$ $\approx -120$ $\mathrm{%
{{}^\circ}%
C}$). For our PDMS, the frequency for glass transition at room temperature
is about $10^{8}$ $\mathrm{Hz}$. A very rough estimate of the exciting
frequency at the level of micro-contacts is $\sim v_{\mathrm{c}}/a$, where $%
v_{\mathrm{c}}=-da/dt$ is the crack tip velocity and $a$ the radius of
micro-spots. As $v_{\mathrm{c}}$ is estimated $<10^{-4}$ \textrm{m/s},
assuming $a$ takes on average values of the order of $10^{-5}$ \textrm{m},
we have exciting frequency less than $10$ $\mathrm{Hz}$. Such values confirm
that viscoelastic losses are negligible within the bulk of micro-asperity
contacts as we are moving inside the rubbery region (or at most on the
border with the transition region).

\subsection{Loading phase}

Numerical simulations of the loading phase are performed by using a discrete
version of the FT multiasperity model \cite{FT}, which is obtained by
calculating the geometry of each asperity rather than using a statistical
distribution for heights. In the original FT model the contact between two
nominally flat surfaces is studied; here, we take into account the spherical
shape of the indenter. 

Each micro-asperity behaves as an isolated contact punch and, according to
the JKR\ formalism \cite{JKR}, the contact load $F$ and approach $\delta $
are given by%
\begin{eqnarray}
F &=&\frac{4}{3}\frac{E^{\ast }a^{3}}{R}-\sqrt{8\pi E^{\ast }\Delta \gamma
_{0}a^{3}}  \label{Fjkr} \\
\delta &=&\frac{a^{2}}{R}-\sqrt{\frac{2\pi a\Delta \gamma _{0}}{E^{\ast }}}%
\text{,}  \label{djkr}
\end{eqnarray}%
where $a$ and $R$ are the contact radius and radius of curvature.

The total contact area and total applied load are obtained by summing up the
contributions of each contacting asperity. For each set of surface and
contact parameters $5$ numerical realizations of the surfaces were
considered.

In this model, the lateral interaction between asperities is not taken into
account as elastic coupling can be reasonably neglected for the considered
surfaces (see Ref. \cite{ACITO2019,YASHIMA2015}). However, we stress that
lateral interactions is instead of crucial importance when the surfaces are
characterized by roughness distributed on several length scales (see, for
example, \cite{Violmultiasperity,ViolanoJKR}).

\subsection{Unloading phase}

Simulations of the unloading phase are performed on the base of the solution
proposed by Muller \cite{Muller1999} and just exploited in Ref. \cite%
{ViolAIAS2019}. Muller showed that the detachment process of a rigid sphere
from a viscoelastic half-space can be described by a two-parameter
differential equation, which in dimensionless form writes%
\begin{equation}
\frac{d\bar{a}}{d\bar{\delta}}=\left[ \frac{\Delta \gamma _{\mathrm{0}}}{%
RE^{\ast }}\right] ^{1/3}\cdot \frac{1}{\beta }\left[ \bar{a}^{3}\left( 1-%
\frac{\bar{\delta}}{3\bar{a}^{2}}\right) ^{2}-\frac{4}{9}\right] ^{1/n}\text{%
,}  \label{eqmuller}
\end{equation}
where $\bar{a}=a/a_{0}$ and $\bar{\delta}=\delta /\delta _{0}$, with $%
a_{0}=3R\left[ \pi \Delta \gamma _{0}/(6E^{\ast }R)\right] ^{1/3}$ and $%
\delta _{0}=3R\left[ \pi \Delta \gamma _{0}/(6E^{\ast }R)\right] ^{2/3}$.
The parameter $\beta $ is given by 
\begin{equation}
\beta =\left( \frac{6}{\pi }\right) ^{1/3}\left( \frac{4}{9}c\right) ^{1/n}V%
\text{.}  \label{beta}
\end{equation}%
being $c$ and $n$ characteristic constants of the material.

This model bases on two main assumptions: i) viscous effects are localized
at the edge of the contact line; ii) detachment occurs under constant
pull-off rate conditions.

In eq. (\ref{eqmuller}), the parameter $\beta $ is proportional to the
pull-off rate $V=-d\delta /dt$, while $n$ may be determined experimentally.
In general, $n$ ranges from $0.1$ to $0.8$ \cite{Muller1999}.
The initial value $\bar{a}(\bar{\delta})$ to solve eq. (\ref{eqmuller}) is
returned by the classical JKR equations, which can be rewritten in
dimensionless form as%
\begin{eqnarray}
\bar{a} &=&\left\{ \frac{1}{2}\left[ 1+\left( 1+\bar{F}\right) ^{1/2}\right]
\right\} ^{2/3}  \label{ain} \\
\bar{\delta} &=&\left( \bar{a}^{2}+\frac{\bar{F}}{2\bar{a}}\right)
\label{din}
\end{eqnarray}%
where $\bar{F}=F/F_{0}$ and $F_{0}=1.5\pi R\Delta \gamma _{0}$. The
dimensionless contact load is then calculated by%
\begin{equation}
\bar{F}=2\bar{a}\left( \bar{\delta}-\bar{a}^{2}\right) \text{.}
\label{Fmuller}
\end{equation}

The FT discrete model returns the value $F_{i}$ of each contacting asperity,
being $F_{i}$ the load reached at the end of the loading process; eqs. (\ref%
{ain}) and (\ref{din}) can hence be used to calculate the values of the
contact radius $a_{i}$ and penetration $\delta _{i}$ at the beginning of the
unloading phase.

In the experiments, unloading is performed by reducing the
mean penetration $\Delta $ of the indenter in the rough surface at controlled displacement rate. A
micro-asperity is assumed to jump out of contact when a critical jump-off
distance is reached. Contrary to the JKR theory, jump out of contact occurs
always at zero contact area.

\section{Results and Discussion}

In order to apply Muller's model for each detaching micro-asperity, we have
to calculate the parameters $n$ and $c$ of eqs. (\ref{eqmuller}-\ref{beta}).
As shown in Ref. \cite{Parte1}, both parameters are scale independent. Their
value can be hence obtained from contact tests on smooth PDMS by fitting the
experimental data with the classical equation of Gent \&\ Schultz \cite%
{GS1972}, relating the energy release rate $G$ to the viscoelastic losses at
the crack tip%
\begin{equation}
G=\Delta \gamma _{0}[1+c\cdot v_{\mathrm{c}}{}^{n}]\text{.}  \label{G2}
\end{equation}

In particular, as shown in Fig. 6 of Ref. \cite{Parte1}, the best fit is
obtained using $c=31$ and $n=0.25$. Notice that, the dimensionality of $c$ is the inverse velocity unit in power of $n$.
During unloading, experiments have been performed at three
different driving velocities $V$ ($V=0.02,$ $0.002,$ $0.0002$ \textrm{mm/s}). Due to the compliance of the cantilever beam, the actual velocity $V_{\textrm{act}}$ of the indenter is lower than the imposed value \cite{Parte1}. The values of the actual velocity $V_{\textrm{act}}$ and the parameter $\beta $ used in Muller's model
are $V_{\textrm{act}}=0.8V$ and $\beta \sim 679,$ $67.9,$ $6.79$ \textrm{mm/s}.

\begin{figure}[tbp]
\begin{center}
\includegraphics[width=15.0cm]{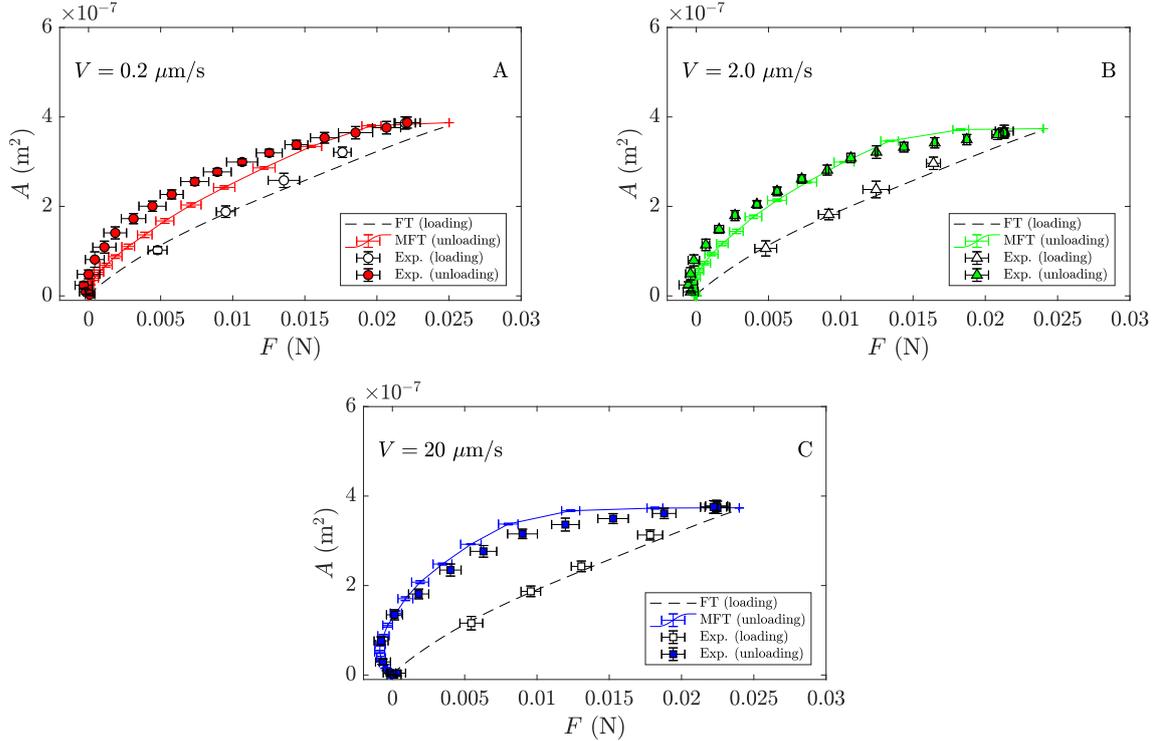}
\end{center}
\caption{The real contact area $A$ as a function of the applied load $F$.
Results are obtained on rough patterns with $\protect\sigma =5$ $\mathrm{%
\protect\mu m}$ and for (A) $V=0.2$ $\mathrm{\protect\mu m/s},$ (B) $V=2$ $%
\mathrm{\protect\mu m/s,}$ and (C) $V=20$ $\mathrm{\protect\mu m/s}$. In all
cases unloading starts after that the load $F_{\max }=0.025$ \textrm{N} is
reached. Experimental data are denoted with markers (open and closed symbols
refer to loading and unloading, respectively). Error bars denote the
standard deviation on $6$ different realizations. Lines refer to numerical
predictions obtained by FT model (loading) and MFT model (unloading).}
\label{ca_V123_sigma5}
\end{figure}

Figs. \ref{ca_V123_sigma5}A-C show the true contact area $A$ as a function
of the applied force $F$ for three values of the detachment rate ($V=0.2,$ $%
2,$ $20$ $\mathrm{\mu m/s}$). Experimental tests are performed on rough
patterns with asperities heights normally distributed with standard
deviation $\sigma =5$ $\mathrm{\mu m}$.

Open and closed symbols refer to loading and unloading experimental data,
respectively. Vertical and horizontal error bars show scattering of results
obtained on $6$ different experiments, corresponding to $6$ contact
realizations. Numerical predictions of the loading phase are plotted with
black dashed line, while solid lines are used for the unloading phase.

Profilometry measurements showed that the vapor treatment used to
smoothening the PMMA mold did not induce any change in the standard
deviation of the asperity height distribution. However, it resulted in a
roughly $10\%$ increase in the radius of curvature of some asperities. For
this reason, contact simulations have been carried out on $5$ numerically
generated surfaces with random distributions of asperities radius of
curvature. Specifically, the radius of curvature has been assumed ranging
from $100$ $\mathrm{%
\mu
m}$ to $110$ $\mathrm{%
\mu
m}$, with an average value of $105$ $\mathrm{%
\mu
m}$. The dispersion of numerical results is shown by the error bars on the
solid lines.

A good agreement is found between experimental data and numerical
predictions. The pull-off process is strongly influenced by the detachment
rate $V$ and, as expected, the pull-off force is enhanced by increasing $V$.

\begin{figure}[tbp]
\begin{center}
\includegraphics[width=15.0cm]{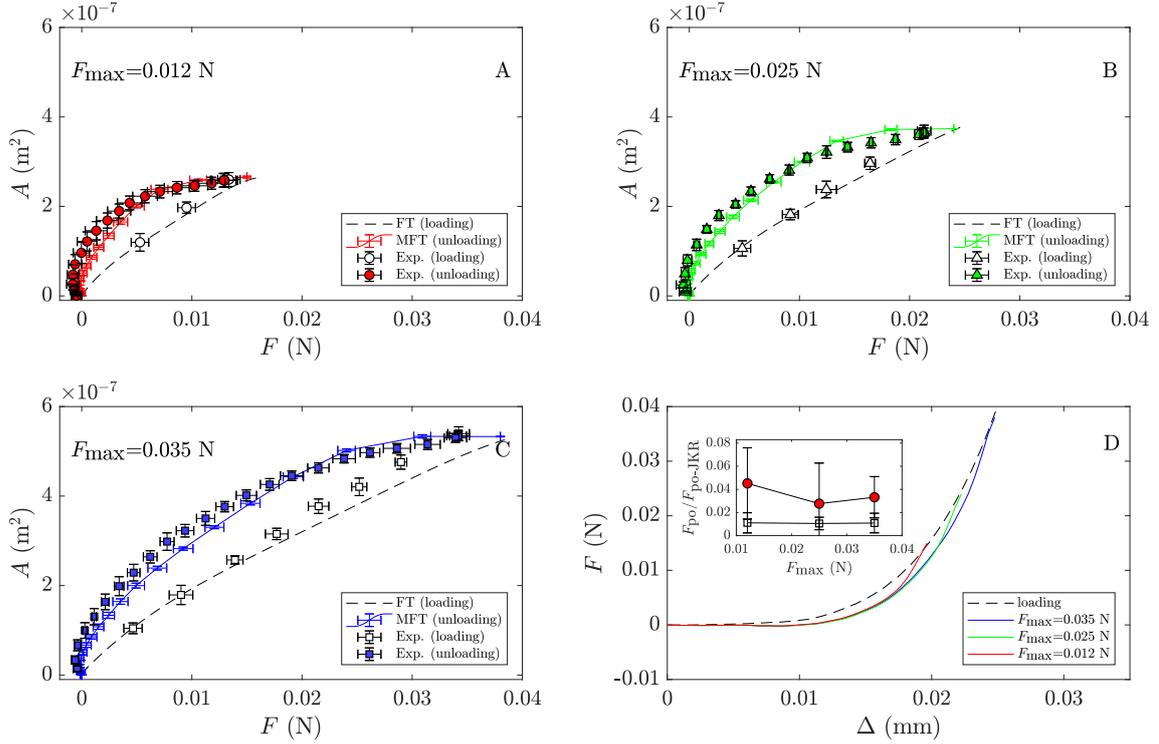}
\end{center}
\caption{The real contact area $A$ as a function of the applied load $F$.
Results are obtained on rough patterns with $\protect\sigma =5$ $\mathrm{%
\protect\mu m}$ and for $V=2.0$ $\mathrm{\protect\mu m/s}$. Moreover,
different values\ of the maximum applied preload are considered: (A) $%
F_{\max }=0.012$ $\mathrm{N}$, (B) $F_{\max }=0.025$ $\mathrm{N}$, and (C) $%
F_{\max }=0.012$ $\mathrm{N}$. Experimental data are denoted with markers
(open and closed symbols refer to loading and unloading, respectively).
Error bars denote the standard deviation on $6$ different realizations.
Lines refer to numerical predictions, obtained by FT model (loading) and MFT
model (unloading). D) The load $F$ as a function of the rigid displacement $%
\Delta $ of the indenter. Lines refer to numerical predictions. In the inset, the normalized pull-off force $F_{\mathrm{po}}$/ $F_{\mathrm{%
po-JKR}}$ as a function of $F_{max}$; red circles and empty squares denote experimental and numerical data, respectively.}
\label{ca_F123_sigma5}
\end{figure}

Figs. \ref{ca_F123_sigma5}A-C show the effect of the maximum applied preload 
$F_{\max }$\ on the contact area vs. applied load relation. Experimental
tests are performed on rough patterns with $\sigma =5$ $\mathrm{\mu m}$, for a fixed unloading detachment rate $V=2$ $\mathrm{\mu m/s}$ and for $F_{\textrm{max}}=0.012, 0.025, 0.035$ N. Once again
we observe a quite good agreement with numerical predictions.

Fig. \ref{ca_F123_sigma5}D shows the variation of the load $F\ $with the
rigid displacement $\Delta $ as predicted by the numerical model. As shown in the inset, the pull-off force is found to be almost independent of the point at which
unloading starts.
Increasing $F_{\max }$ of a factor $\sim 3$
leads to a $74\%$ increase in the energy loss for adhesion hysteresis. In
the contact of smooth elastic bodies, JKR\ theory predicts the pull-off
force to be independent on the magnification of the preload $F_{\max }$.
This result is still valid for viscoelastic media. In the case of rough
contacts, it is not completely understood how surface roughness affects the
dependence of the pull-off force on the maximum applied load. Recent
experimental investigations \cite{dorogin2017} show an increase in the
pull-off force with $F_{\max }$. However, such findings disagree with tests
on rough PDMS performed by Kesari et al. \cite{kesari2010}, which found a
little, almost negligible, enhancement of the pull-off force with $F_{\max }$
in agreement with the Greenwood's statement: "...\textit{in a number of
calculations the pull-off force has proved to be almost, or completely,
independent of the point at which unloading starts, although the initial
parts of the curve certainly do differ}." (Ref. \cite{greenwood2017}).

\begin{figure}[tbp]
\begin{center}
\includegraphics[width=15.0cm]{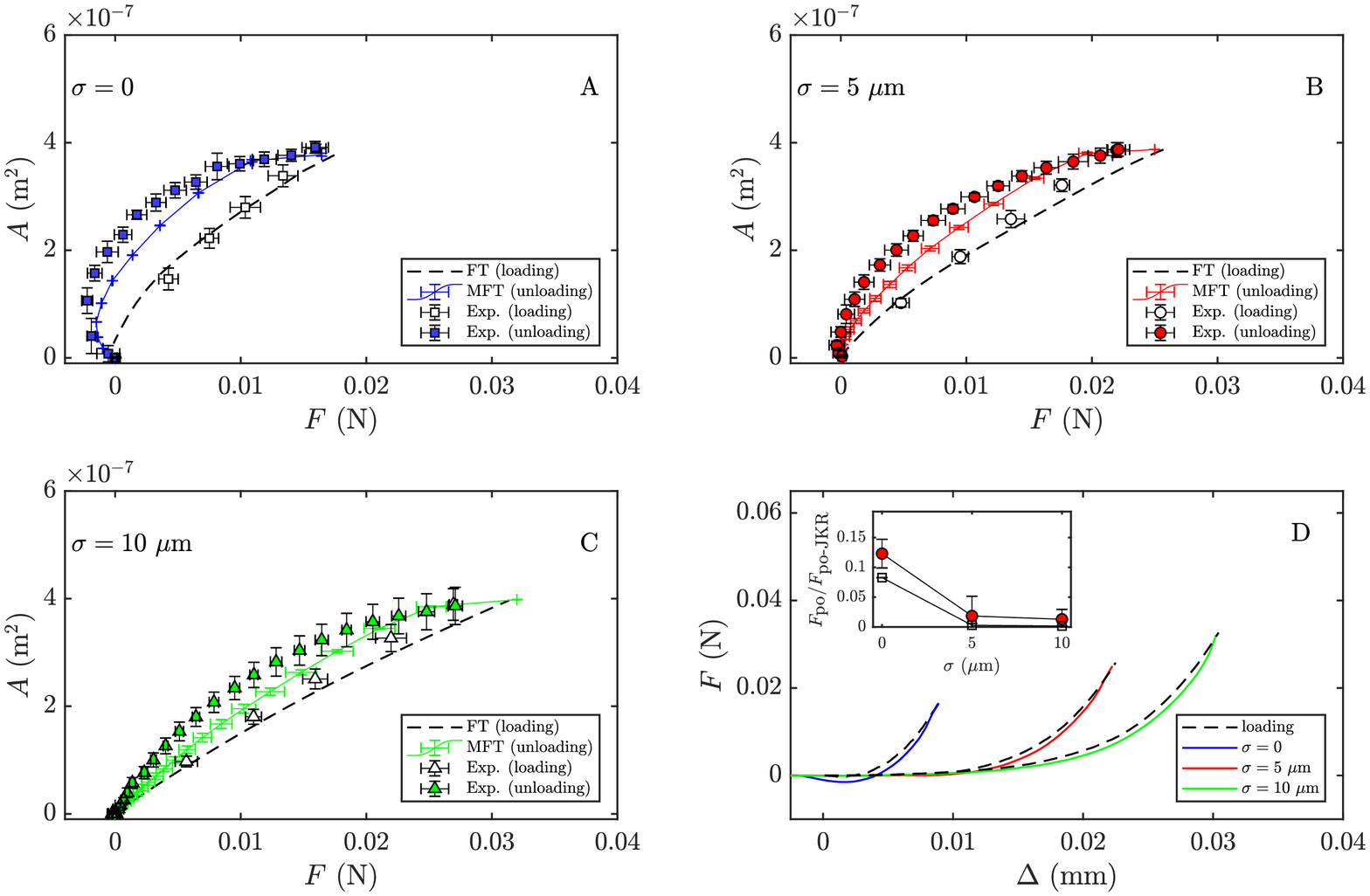}
\end{center}
\caption{ The real contact area $A$ as a function of the applied load $F$.
Results are obtained on rough patterns with $\protect\sigma =0$ (A), $%
\protect\sigma =5$ $\mathrm{\protect\mu m}$ (B), $\protect\sigma =10$ $%
\mathrm{\protect\mu m}$ (C), and for $V=0.2$ $\mathrm{\protect\mu m/s}$.
Experimental data are denoted with markers (open and closed symbols refer to
loading and unloading, respectively). Error bars denote the standard
deviation on $6$ different contact realizations. Lines refer to numerical
predictions, obtained by the FT (loading) and MFT (unloading) models. D) The
load $F$ as a function of the rigid displacement $\Delta $ of the indenter.
Lines refer to numerical predictions. In the inset, the normalized pull-off force $F_{\mathrm{po}}$/ $F_{\mathrm{%
po-JKR}}$ as a function of $\sigma$; red circles and empty squares denote experimental and numerical data, respectively.}
\label{ca_s123_V1}
\end{figure}

Figs. \ref{ca_s123_V1}A-C show the area-load curves during loading and
unloading phases for three values of $\sigma $ ($\sigma =0,$ $5,$ $10$ $%
\mathrm{\mu m}$). Unloading tests were performed at fixed detachment rate $%
V=0.2$ $\mathrm{\mu m/s}$. The agreement between experiments and numerical
data is generally very good.
Figs. \ref{ca_s123_V1}D shows how the $F-{\Delta}$ relation modifies with the roughness amplitude. We notice that the pull-off force vanishes when increasing $\sigma $, in agreement with the findings of Fuller \&
Tabor \cite{FT} and more recent studies \cite{ACITO2019}.

However, recent experimental works find that roughness may increase
adhesion. Indeed, in contact tests between a spherical tip and PMDS with
nanometer scale roughness, Kesari et al. \cite{kesari2010} found an `optimal
roughness' value, which maximizes the hysteretic energy loss.

Moreover, Dalvi et al. \cite{dalvi2019} performed adhesion measurements on
soft elastic PDMS hemispheres in contact with polycrystalline diamond rough
substrates. Loading-unloading tests were conducted under quasi-static
conditions, i.e. at very small values of the driving velocity of the
indenter. The authors found that, even with negligible viscous effects,
adhesion hysteresis still occurs because of a roughness-induced increase in
contact area, in agreement with the predictions of the model of Persson \&
Tosatti \cite{PT2001}. Moreover, Greenwood \cite{greenwood2017} extended the
FT model to the unloading phase and found adhesion hysteresis occurs also in
absence of viscous effects because of elastic instabilities.

\subsection{Hysteretic dissipation}

\begin{figure}[tbp]
\begin{center}
\includegraphics[width=16.0cm]{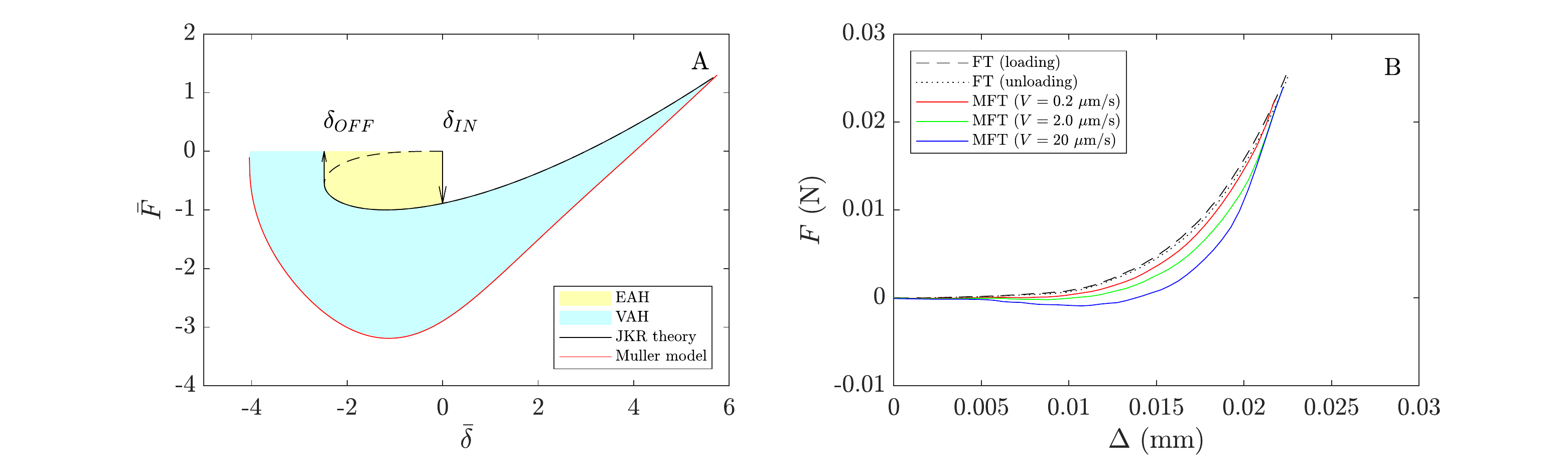}
\end{center}
\caption{A) The force-approach relation for smooth elastic spheres, as
predicted by the JKR\ and Muller models. The yellow and cyan areas represent
the energy loss due to elastic instabilities (EAH) and viscoelastic
dissipation (VAH), respectively. B) The applied load $F$ as a function of
the rigid displacement $\Delta $ of the indenter as predicted by the
theoretical model. The loading path is plotted with black dashed lines (FT
model); the unloading elastic path is plotted with black dotted line (FT
model); the unloading viscoelastic path is plotted with colored solid lines
(MFT model). Results are obtained on rough patterns with $\protect\sigma =5$ 
$\mathrm{\protect\mu m}$ and for $V=0.2,$ $2.0,$ $20$ $\mathrm{\protect\mu %
m/s}$ (red, green and blue).}
\label{s5_isteresi_1}
\end{figure}

For a rigid smooth sphere approaching a flat compliant substrate, the
loading-unloading path predicted by JKR theory and Muller model are shown in
Fig. \ref{s5_isteresi_1}A. In JKR theory (black line), loading and unloading
curves overlap and hysteretic energy loss (yellow area) is related to the
elastic instabilities due to the different penetrations at which jump to
contact ($\delta _{IN}$) and detachment ($\delta _{OFF}$) occur. In
practical cases, such energy loss is usually negligible. On the contrary, in
presence of viscous effects, the unloading path (red line) is rate-dependent
and the hysteretic energy loss is much larger.

When a distribution of micro-asperities is textured on the substrate, the
loading-unloading curves modify as shown in Fig. \ref{s5_isteresi_1}B.
Numerical predictions are shown for $\sigma =5$ $\mathrm{\mu m}$. Negligible
hysteresis occurs when viscous effects are neglected, demonstrating that the
origin of adhesion hysteresis in our experiments is strongly related to
viscoelasticity. However, such result is related to have considered a single
length-scale roughness. In reality, for roughness characterized by several
length scales, we expect larger elastic adhesion hysteresis as a result of a
more significant roughness-induced increase in contact area, as shown
experimentally in Ref. \cite{dalvi2019}. Furthermore, roughness with several
length scales may enhance localized phenomena of elastic instability Ref. 
\cite{carbone2015}.

\begin{figure}[tbp]
\begin{center}
\includegraphics[width=16.0cm]{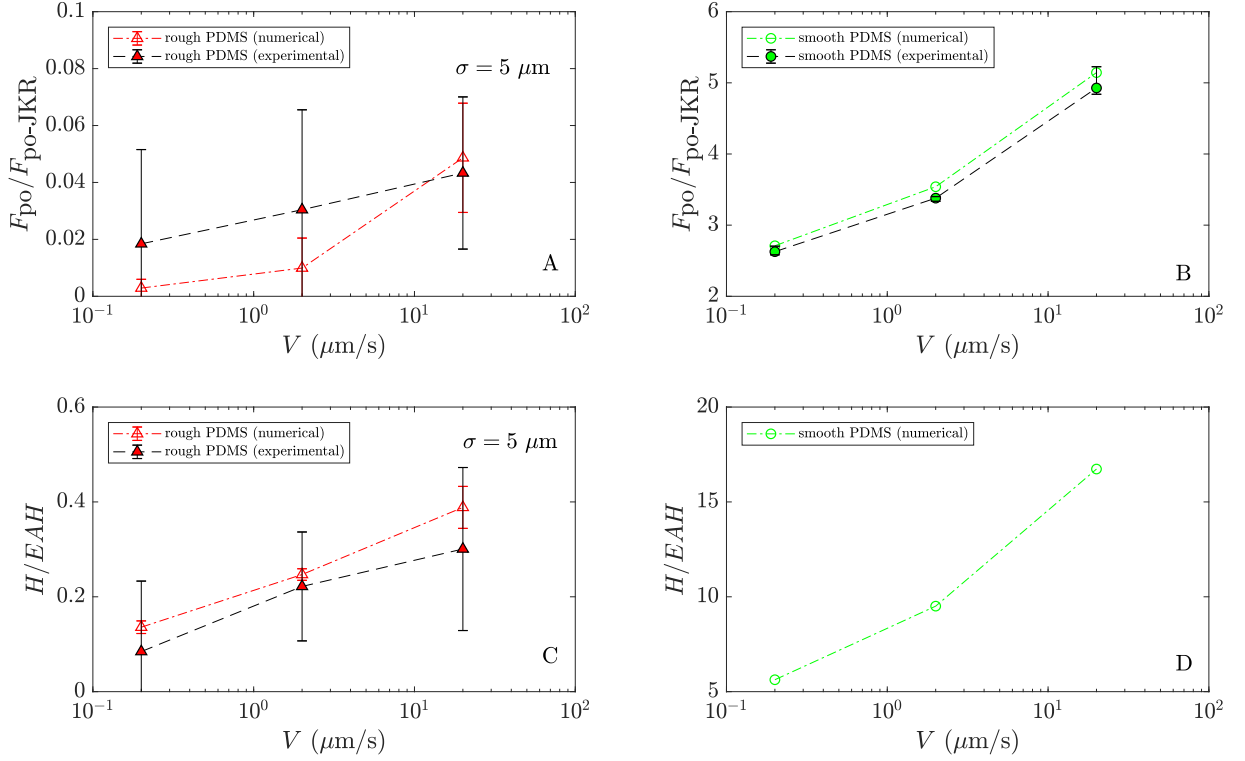}
\end{center}
\caption{A) The normalized pull-off force $\ F_{\mathrm{po}}/F_{\mathrm{%
po-JKR}}$ for rough PDMS samples. Numerical (open triangles) and
experimental (filled triangles) data are shown for $\protect\sigma =5$ $%
\mathrm{\protect\mu m}$ and $V=0.2,$ $2.0,$ $20$ $\mathrm{\protect\mu m/s}$.
B) The normalized pull-off force calculated for smooth PDMS samples. C) The
normalized hysteretic loss$\ H/EAH$ obtained on rough PDMS samples. $EAH$ is
the energy dissipated as a result of the elastic instabilities predicted by
the JKR\ theory for smooth contacts. D) The normalized hysteretic loss
calculated for smooth PDMS samples.}
\label{s5_isteresi_2}
\end{figure}

In our experiments, surface roughness reduces the pull-off force by a factor 
$\sim 120$ as shown in Fig. \ref{s5_isteresi_2}A where, for $\sigma =5$ $%
\mathrm{\mu m}$, the pull-off force $F_{\mathrm{po}}$ estimated in the tests
on rough PDMS is plotted in terms of the driving velocity $V$. Notice $F_{%
\mathrm{po}}$ is normalized with respect to the pull-off force $F_{\mathrm{%
po-JKR}}=1.5\pi R\Delta \gamma _{0}$ evaluated on a smooth sample and
neglecting viscous dissipation. In Fig. \ref{s5_isteresi_2}B similar curves
are given for the case of smooth sample. Results shown an increase in $F_{%
\mathrm{po}}$ with $V$ and a good agreement between experimental and
numerical predictions (both at macro and microscale).

Tiwari et al. \cite{tiwary2017} performing\ contact experiments on stiff
PDMS\ ($E^{\ast }\sim 2$ \textrm{MPa}) observed that roughness leads to a
decrease in $\Delta \gamma _{\mathrm{eff}}$ of a factor $\sim 700$. However,
for very soft PDMS ($E^{\ast }\sim 0.02$ \textrm{MPa}) they found an
enhancement of $\Delta \gamma _{\mathrm{eff}}$. The latter was observed
close to full-contact conditions, because of roughness-induced increase in
the real contact area. In our experiments we are far from full-contact
conditions and this explains why we do not observe any increase in pull-off
force moving from smooth to rough contact.

Moreover, pull-off force and effective surface energy are expected to be
influenced by viscous dissipation. Under the assumption of viscous effects
located near the crack tip only, the dissipation due to the crack
propagation mechanism is proportional to the length $L$ of the boundary line
between contact and non-contact areas \cite{krick2012}. Under a compressive
load $F=0.02$ \textrm{N}, we find that the total perimeter $L_{m}$ of
micro-contact spots is of the same order of magnitude of the perimeter $L_{M}
$ of the smooth macro-spot ($L_{m}\approx 2L_{M}$ ). For this reason we
could expect a slight increase in $\Delta \gamma _{\mathrm{eff}}$; however,
the contact area $A$ reduces more significantly (by a factor of $\sim 31$)
as a result of the surface roughness. Such reduction is predominant and
therefore governs the variation of $\Delta \gamma _{\mathrm{eff}}$ in
agreement with that observed in Ref. \cite{Isra2000}: "...\textit{the
effective adhesion force or surface energy per unit area can be very low -
often many orders of magnitude below the value for two molecularly smooth
surfaces - and is determined by a few isolated asperity contacts of low
radii of curvature}".

The surface roughness is expected to have the same effect on hysteretic
losses in agreement with previous observations given in Szoszkiewicz et al. 
\cite{Szo2006}, who performing measurements of adhesion hysteresis at
nanometer and micrometer length scales on mica, calcite, and metallic
samples, found that hysteretic losses decrease of two orders of magnitude
moving from micro to nano sized particles.

In this regard, the hysteretic loss occuring during a loading-unloading
cycle is proportional to the area $H$\ enclosed between the loading and
unloading curves shown in Fig. \ref{s5_isteresi_1}B. Its variation with the
driving velocity $V$ is plotted in Fig. \ref{s5_isteresi_2}C, where $H$ is
normalized with respect to the value calculated for a smooth sample in
absence of viscous dissipation. In Fig. \ref{s5_isteresi_2}D the same plot
is given for the case of smooth substrate. As predicted, a great reduction
in adhesion hysteresis is registered when experiments (and numerical
simulations) are carried out on rough samples as a result of the decrease in
the real contact area. Moreover, the hysteretic losses increase with the
driving velocity which is clearly related to the crack tip velocity $v_{%
\mathrm{c}}$ in a way that depends on the crack tip process zone.

\section{Conclusions}

In this work, we performed experimental and numerical investigations on the
adhesive normal contact between a spherical indenter and viscoelastic rough
substrates finding a quite good agreement.

The proposed numerical model makes use of a discrete version of the FT
multiasperity model to describe the loading phase and the solution proposed
by Muller \cite{Muller1999} to characterize the unloading one. Moreover, as
we can also perform the unloading process neglecting viscous effects, we can
distinguish the hysteretic energy loss due to viscous dissipation from that
due to the roughness-induced increase in contact area.

Both experimental and numerical results show that adhesion is strongly
enhanced by increasing the detachment rate and decreasing the rms roughness
amplitude, while the pull-off force is negligibly affected by the maximum
applied load. This last trend is in agreement with previous studies \cite%
{greenwood2017,kesari2010}, but it seems to contradict the increase in
pull-off force observed in Ref. \cite{dorogin2017}. However, we have to
pointed out that our calculations are performed on simplified rough surfaces
characterized by a single length scale.

Furthermore, we find that the increase in the effective adhesion energy with
the crack tip velocity is independent on the size of the radius of
curvature, depending on the viscoelastic properties of bodies exclusively.
Interestingly, the increase in the perimeter of the contact line observed on
rough samples negligibly affects the effective adhesion energy $\Delta
\gamma _{\mathrm{eff}}$, which is instead strongly affected by the reduction
in contact area that, in our experiments, is therefore the parameter
governing the change in surface energy observed when moving from smooth to
rough samples.

The present study has limitations related to the simplistic description of
the surface roughness, which usually presents fractal features. However, we
believe it can be useful in clarifying some key points in the adhesion
hysteresis of rough soft media ("\textit{Asperity approach leads naturally
to an understanding of the difference between loading and unloading, and so
to why there is hysteresis}", Ref. \cite{greenwood2017}).

\end{document}